\documentclass[preprint,superscriptaddress,preprintnumbers,amsmath,amssymb]{revtex4}

\usepackage{graphicx}
\usepackage{dcolumn}
\usepackage{bm}
\usepackage{color}

\begin{document}

\preprint{}

\title{Weak temporal signals can synchronize and accelerate the transition dynamics of biopolymers under tension}


\author{Won Kyu Kim}
 \affiliation{Department of Physics and POSTECH Center for Theoretical Physics, Pohang University of Science and Technology (POSTECH), Pohang, 790-784, Republic of Korea}
\author{Changbong Hyeon}
 \thanks{Corresponding author} 
  \email{hyeoncb@kias.re.kr}
 \affiliation{School of Computational Sciences, Korea Institute for Advanced Study, Seoul 130-722, Republic of Korea} 
\author{Wokyung Sung}
  \thanks{Corresponding author} 
  \email{wsung@postech.ac.kr}
 \affiliation{Department of Physics and POSTECH Center for Theoretical Physics, Pohang University of Science and Technology (POSTECH), Pohang, 790-784, Republic of Korea}

\date{\today}

\begin{abstract}
In addition to thermal noise, which is essential to promote conformational transitions in biopolymers, cellular environment is replete with a spectrum of athermal fluctuations that are produced from a plethora of active processes. To understand the effect of athermal noise on biological processes, we studied how a small oscillatory force affects the thermally induced folding and unfolding transition of an RNA hairpin, whose response to constant tension had been investigated extensively in both theory and experiments. Strikingly, our molecular simulations performed under overdamped condition show that even at a high (low) tension that renders the hairpin (un)folding improbable,
a weak external oscillatory force at a certain frequency can synchronously enhance the transition dynamics of RNA hairpin and increase the mean transition rate. Furthermore, the RNA dynamics can still discriminate a signal with resonance frequency even when the signal is mixed among other signals with nonresonant frequencies. In fact, our computational demonstration of thermally induced resonance in RNA hairpin dynamics is a direct realization of the phenomena called stochastic resonance (SR) and resonant activation (RA). Our study, amenable to experimental tests using optical tweezers, is of great significance to the folding of biopolymers in vivo that are subject to the broad spectrum of cellular noises.
\end{abstract}




\maketitle

The cell is a highly crowded and dynamic place with millions of biomolecules and their interactions \cite{AlbertsBook}.
Only recently the notion that the self-assembly and function of biomolecules in vivo could be very different from those in isolation has been highlighted in terms of macromolecular crowding and confinement \cite{Zhou2008ARB,dhar10PNAS,Kim11PRL}.
However, the implication from dynamics in vivo should not necessarily be limited to the confinement or crowding effect.
Besides the thermal noise, which is essential for activated internal motions and folding dynamics of biomolecules, of particular note are the nonequilibrium athermal signals produced from a plethora of active processes and collective actions among cellular constituents.
For example, ribosome translocate along mRNA strand less than every 0.1 s per codon \cite{Wen08Nature};
Molecular motor, kinesin, takes step along the microtubule approximately every $10$ ms \cite{Visscher99Nature};
Collective beating of cilia occurs approximately every $1$ min \cite{sanchez2011cilia}.
How these nonequilibrium signals or noises affect fundamental processes in biology, particularly the folding dynamics of biopolymers \cite{OnuchicCOSB04,Thirumalai10ARB}, is the main concern of this paper.
To address this issue, we study the effect of a weak oscillatory force on the transition dynamics of stretched RNA hairpins as a test case.

In recent years a combination of experiments, theoretical arguments and simulations have been used to extensively study the conformational dynamics of biopolymers under mechanical control \cite{Bustamante94SCI,Bustamante01Sci,Bustamante03Science,Block06Science,Greenleaf08Science,Marqusee05Science,stigler2011complex,Ritort06JPHYS,HyeonBJ07,Hyeon08PNAS,HyeonMorrison09PNAS}.
Among others, paramount precision of optical tweezers, which can both exert piconewton force to molecules and probe structural changes in nm scale, not only allowed us to quantify thermodynamic and kinetic properties of the molecule at single molecule level \cite{Bustamante01Sci,Block06Science}, but also provided unprecedented chances to manipulate the process of life \cite{Visscher99Nature,Strick00Nature,Smith01Nature,Purohit03PNAS,BlockNature03}.
To emulate nonequilibrium athermal noise in vivo and study its effect on the conformational dynamics of biopolymers, high precision force instruments seem to hold good promise; yet, no experiment to date has explored this issue.

Given that the dynamics of proteins and nucleic acids occur at overdamped fluctuating media, it may be surprising to learn that conformational transitions of biopolymers could become enhanced or synchronized to an externally exerted weak oscillatory signal.
However, according to the notion of stochastic resonance (SR), sensitivity of a system to external signal can be improved under thermal fluctuation, so that the system can discriminate the weak information carrying signals \cite{gammaitoni1998stochastic}.
Theoretically, it has been shown for a single Brownian particle in a bistable potential that synchronization of barrier crossing events to an externally driven oscillation occurs not only at an optimal noise strength but also at resonance frequency \cite{gammaitoni1998stochastic,sr1}. 
Furthermore, at another oscillation frequency that differs from the frequency of synchronization, the mean first passage time over the barrier is minimized, which is called the resonant activation (RA) \cite{doering1992resonant}.
Hence, it is possible to demonstrate the SR and RA in the conformational transition dynamics of biopolymers like RNA hairpins, whose folding landscape is bistable near transition mid-force, by properly tuning the frequency of additive oscillatory force.

Since the pioneering work of Benzi \emph{et al.}
\cite{0305-4470-14-11-006}, there have been efforts to extend the idea of SR to a vast range of physical and biological phenomena \cite{gammaitoni1998stochastic,Frauenfelder99BP,srbio}.
However, SR often mentioned
in terms of macroscopic systems such as recurrence of the
ice age cycle and sensory systems for higher order organisms,
especially the latter of which is wired with complex biochemical
networks, is always a subject of debate because the inherent complexity
of a system consisting of a formidable number of variables
often makes the origin of such resonance-like phenomena inconclusive.
Although phenomenological modeling of system dynamics using SR is still possible, there seems to be little convincing proof that the biological systems actually use SR in the cellular level \cite{McDonnell09PLoSComp}.
In this aspect, recent theoretical demonstrations of SR and RA of a polymer chain in a confined space has more direct relevance to the biology \cite{asfaw,jaeoh}.
Here we show that SR and RA can indeed be manifested in self-organization process of biopolymers by using molecular simulation.
Although careful investigation is further required to confirm whether the
SR and RA are actively utilized in the cell for its biological functions,
it should be straightforward to demonstrate SR in the folding
and unfolding dynamics of biopolymers by using single
molecule force experiments. We provide a simple experimental protocol for the demonstration.

\section*{RESULTS AND DISCUSSIONS}

{\bf Bistable transition of P5GA hairpin under constant tension :}
To study the dynamics of biopolymers under tension, we used the self-organized polymer (SOP) model, a new class of versatile coarse-grained model that have been proven powerful in describing diverse problems in biomolecular dynamics at single molecule level, which includes proteins, RNA, and molecular motors \cite{HyeonBJ07,Hyeon08PNAS,HyeonMorrison09PNAS, Hyeon06Structure,Hyeon06PNAS,Hyeon07PNAS2}.
Among the molecules studied using SOP model, the 22-nucleotide P5GA RNA hairpin is the simplest one, whose dynamics was studied under various mechanical conditions such as force clamp and force ramp (See Methods), providing microscopic insights into RNA hairpins in single molecule force experiments \cite{HyeonBJ07,Hyeon08PNAS,HyeonMorrison09PNAS}.
Here, we performed Brownian dynamics simulation of the RNA hairpin under varying tensions (see Methods).
As shown previously \cite{HyeonBJ07,Hyeon08PNAS,HyeonMorrison09PNAS}, the time trajectories that describe the dynamics of molecular extension ($z$) show that the RNA hairpin undergoes thermally-induced hopping transitions when the molecule is held around the transition mid-force $f_m$.
The force-dependent free energy profile $F(z;f)=-k_BT\log{P(z;f)}$ with $T=300$ K, calculated from the distribution of molecular extension $P(z;f)$, becomes bistable at $f=f_m$ ($F(z;f_m)=F_m(z)$) by forming the native basin of attraction (NBA) and unfolded basin of attraction (UBA) at $z_{F}\approx2$ nm and $z_{U}\approx8$ nm, respectively (Fig.~\ref{fig:1}a).
The tension greater (or smaller) than $f_m$, $f>f_m$ (or $f<f_m$) tilts the free energy profile towards the UBA (or NBA).
The force-dependent probability of being in the UBA $P_{U}(f)$ is calculated using $P_U(f)=\int^{z_{M}}_{z_{TS}} dz e^{-\beta F(z;f)}/\int^{z_{M}}_{z_{m}} dz e^{-\beta F(z;f)}$ where $\beta=(k_B T)^{-1}$, $z_{TS}$ is the extension corresponding to the barrier top (transition state), $z_m$ and $z_{M}$ (see Fig.~\ref{fig:1}a) are the minimum and maximum extensions, respectively.
$P_U(f)$, fitted to a two state model, $P_{U}^{*}=[1 + e^{{(f_m - f )\Delta z}/k_B T}]^{-1}$ \cite{Bustamante01Sci}, gives $f_m=14.7$ pN and $\Delta z\approx 6$ nm (Fig.~\ref{fig:1}b).
From the time trajectories, the $f$-dependent mean folding ($\tau_F(f)$) and unfolding time ($\tau_U(f)$) are also calculated and found to ``approximately" obey the Bell-like relation \cite{Bell78SCI} (Fig.~\ref{fig:1}c). It is of note that the deviation from linearity on a log folding time scale is due to moving transition state resulting from the underlying folding landscape that deforms under mechanical force \cite{Hyeon06BJ,Dudko06PRL,Hyeon07JP}.\\

{\bf Oscillation-induced coherent transition of the P5GA hairpin :}
According to theories, the nonlinear and bistable characteristics of the folding landscape of RNA hairpin guarantee the manifestation of SR and RA if an oscillatory driving is externally applied to the molecule just like the SR discussed in the context of Brownian particle on a bistable potential.
However, given that there are many degrees of freedom whose dynamics experiences continuous dissipation over time, it is not obvious whether a small oscillatory force added to the RNA can indeed be sensed along the backbone of biopolymer and renders the folding transition synchronous to the oscillatory force.
For proteins and RNAs where entropy play dominant role in shaping the underlying free energy landscape, the thermal noise strength (temperature), kept at a physiological condition, is not a good variable to vary.
Instead, it would be more appropriate to vary the frequency of oscillatory force.
To this end, we performed the many-body simulations using SOP model by applying a small oscillatory force $\delta f \sin{(\Omega t)}$ with $\delta f=1.4$ pN to the 5'-end of P5GA hairpin at $f=17$ pN and monitor the change in folding-unfolding dynamics, for various driving frequencies $\Omega=2\pi/T_{\Omega}$.

Our many-body simulations using the SOP model of RNA hairpin show that although the probability of being in the NBA at $f=17$ pN is only $\approx 5$ \% so that the folding is improbable, folding transitions occur more frequently and in coherence to a small oscillatory force at the period of $T_{\Omega}=T^{SR}\simeq10.2$ ms (Fig.~\ref{fig:2}c).
Here, it is noteworthy that $T^{SR}$ for our RNA hairpin at $f=17$ pN is similar to the sum of the two transition times in the absence of oscillatory driving, $\tau_{F}(f)+\tau_{U}(f)\approx 10.8$ ms (see Fig.~\ref{fig:1}c at $f=17$ pN), a theoretically well-known SR condition for symmetric bistable potential \cite{gammaitoni1998stochastic}.\\

{\bf Mapping P5GA dynamics onto one-dimensional free energy profile :}
To explore the details of SR and RA in the dynamics of RNA hairpin more extensively, 
we simplified our many-body simulation using SOP model into one-dimensional (1D) simulation by showing that the dynamics of molecular extension $z(t)$ from many-body dynamics can effectively be projected onto the dynamics of a quasi-particle under the 1D free energy profile $F(z;f)$.
By using $f$-dependent free energy profile $F(z;f)=F_m(z)-(f-f_{m})~z$ where the $F_m(z)$ is directly interpolated from the energy profile at $f_{m}=14.7$ pN obtained by the SOP simulation (Fig.~\ref{fig:1}), we modeled an associated Langevin dynamics of the quasi-particle with a position $z(t)$ as follows:
$\gamma\dot{z}=-\partial F(z;f)/\partial z+\xi(t)$.
Here, $\gamma$ is the effective friction coefficient and $\xi(t)$ is the random force that satisfies the fluctuation-dissipation theorem $\langle\xi(t)\xi(t')\rangle=2k_{B}T\gamma\delta(t-t')$ with $\langle\xi(t)\rangle=0$.
Folding and unfolding times, determined from time trajectories $z(t)$ on $F(z;f)$, $\tau^{1D}_{F}(f)$ and $\tau^{1D}_{U}(f)$ are in excellent agreements with those from the SOP model ($\tau_F(f)$ and $\tau_U(f)$) (Fig.~\ref{fig:1}c).
In addition, by imposing reflecting boundary condition at $z_m\approx 1.01$ nm and $z_M\approx 9.7$ nm, which are the minimum and maximum molecular extension, we analytically calculated the mean first passage (Kramers) times  $\tau_{F}^{KR}(f)=\frac{1}{D}\int^{z_U}_{z_F}~{dz} e^{\beta F(z;f)}\int_{z}^{z_M}~{dz'} e^{-\beta F(z';f)}$ for the folding and $\tau_{U}^{KR}(f)=\frac{1}{D}\int^{z_U}_{z_F}~{dz} e^{\beta F(z;f)}\int_{z_m}^{z}~{dz'} e^{-\beta F(z';f)}$ for the unfolding.
We set the effective diffusion constant $D={k_{B}T}/{\gamma}\approx13.7$ $\mathrm{\mu m^2/s}$ to match $\tau_F^{KR}(f)$ ($\tau_U^{KR}(f)$) (solid curves in Fig.~\ref{fig:1}c) to $\tau_F(f)$ ($\tau_U(f)$) at $f=f_{m}$ from the simulation results using SOP model (circles in Fig.~\ref{fig:1}c).

The excellent agreement among differently calculated mean transition times $\tau_F(f)\approx \tau_F^{1D}(f)\approx \tau_F^{KR}(f)$ justifies mapping the many-body dynamics of P5GA hairpin under varying $f$ onto the dynamics on 1D free energy profile described with molecular extension, $z$.
In fact, previous studies showed that $z$ is a good reaction coordinate on which both thermodynamics and kinetics of P5GA hairpin are faithfully described as long as $f$ does not greatly differ from $f_m$ \cite{Hyeon08PNAS,HyeonMorrison09PNAS}.
This enables us to explore SR and RA of the RNA folding-unfolding dynamics in a simpler way without compromising the goal of our study.
To take into account the effect of external oscillatory force, we simply added a term $\delta f \sin(\Omega t)$;
thus, the stochastic equation describing the dynamics of $z(t)$ in the presence of external oscillatory signal reads as follows:
\begin{align}
\gamma\dot{z}=-\frac{\partial F_m(z)}{\partial z}+(f-f_m)+\delta f \sin(\Omega t)+\xi(t).
\label{eq:sv}
\end{align}\\

{\bf Resonant activation in the folding and unfolding transitions of RNA hairpins :}
To investigate the effect of oscillatory driving on transition kinetics,
we calculated the average folding time $\tau_F$ as a function of $T_{\Omega}$ for time trajectories generated using Eq.\ref{eq:sv} at $f=17$ pN for $\delta f=1.4$, 2.0 pN or unfolding time $\tau_U$ at $f=13$ pN for $\delta f=1.0$, 1.4 pN (Fig.~\ref{fig:3}).
We found that the external oscillatory signal accelerates both folding and unfolding dynamics and maximizes the transition rates $($defined by $\tau_F^{-1},\tau_U^{-1})$ at $T_{\Omega}=T^{RA}\approx 2.56$ ms by about 60 \% and 30 \% for both folding and unfolding, compared with those in the absence of the oscillatory driving.
The enhancement of folding (unfolding) rate could still be called RA, following the theoretical studies on symmetric bistable potential \cite{doering1992resonant}.

The presence of optimal condition of $T_{\Omega}=T^{RA}$ that minimizes the transition time indicates that there should be a matching condition for two frequencies between conformational fluctuation in folding landscape, which is induced by thermal noise, and the oscillation of input signal.
For $T_{\Omega}\ll T^{RA}\lesssim\tau$, net driving force provided by the fast oscillating force is essentially zero; $\int^{\tau+t_0}_{t_0}\delta f\sin{(2\pi t/T_{\Omega})}dt\approx \int^{\infty}_0T_{\Omega}\delta f\sin{(2\pi s)}ds\approx 0$.
In contrast, when $T^{RA}\ll T_{\Omega}\lesssim \tau$, the molecule experiences external tension that slowly varies between $f-\delta f$ and $f+\delta f$ during the transition time $\tau$ depending on the time interval $t_0<t<t_0+\tau$.
As a result, for $T_{\Omega}\approx 100$ ms, $\tau_F$ at  $f=17$ pN and $\delta f=1.4$ pN varies between 2 ms and 30 ms (Fig.~\ref{fig:1}c), and the corresponding mean folding time becomes $\overline{\tau}_F=\frac{1}{\tau}\int^{t_0+\tau}_{t_0}P_{F}(f(t))\tau_F(f(t))dt \approx 9$ ms as shown in Fig.~\ref{fig:3}a, which is greater than the $\tau_{F}\approx8$ ms at $T_{\Omega}=T^{RA}$ \cite{spagnolo}.

The extent of RA varies with the amplitude of oscillatory force ($\delta f$) (compare circles and squares in Fig.~\ref{fig:3}).
It is conceivable that $\delta f\times \delta z^{\ddagger}_F$ where $\delta z^{\ddagger}_F=z_U-z_{TS}\approx 4$ nm at $f=17$ pN (Fig.~\ref{fig:1}) provides extra work to rectify the random fluctuation of molecule in UBA. Therefore, the ratio of $\delta f\times \delta z^{\ddagger}_F$ to free energy barrier $\Delta F^{\ddagger}$ from the UBA to NBA, i.e., $\delta f\times \delta z^{\ddagger}_F/\Delta F^{\ddagger}$ should determine the enhancement of folding rate at the RA condition.\\

{\bf Stochastic resonance in the refolding dynamics of RNA :}
Folding time distributions $p_{fold}(t)$ (Fig.~\ref{fig:4}b) calculated from time trajectories (Fig.~\ref{fig:4}a) are useful measure for characterizing SR.
At fast driving $T_{\Omega}=0.16$ ms (Fig.~\ref{fig:4}b-1), $p_{fold}(t)$ decays exponentially with a decay time 10.5 ms, which is consistent with the folding time in the absence of driving, $\tau_F^o$ (see Fig.~\ref{fig:1}c).
But at $T_{\Omega}=10.24$ ms near the resonant condition, $p_{fold}(t)$ displays multiple peaks whose periodicity  corresponds to the driving period of input signal, which is approximately equal to $\tau_F^o$ (Fig.~\ref{fig:4}b-2).
The multiple peaks are superimposed on the above-mentioned exponentially decaying contribution (Fig.~\ref{fig:4}b-1) \cite{gammaitoni1998stochastic}.
At a slow driving $T_{\Omega}=40.96$ ms (Fig.~\ref{fig:4}b-3), the $p_{fold}(t)$ has a broad and small peak around 25 ms and shows only a weak signature of coherence.

The degree of synchronization is quantified by calculating the $P_1$ value corresponding to the area under the first peak of $p_{fold}(t)$, which is defined as $P_{1}=\int_{\tau_{min_1}}^{\tau_{min_2}} dt p_{fold}(t)$ where the integration is done over a time interval between $\tau_{min_1}$ and $\tau_{min_2}$ that defines the first peak in the $p_{fold}(t)$ \cite{gammaitoni1998stochastic,sr1,srra}.
For example, for $T_{\Omega}=0.16$ ms,
the $p_{fold}(t)$ exponentially decreases with no peak, thus $P_{1}=0$.
In contrast, $p_{fold}(t)$s with $T_{\Omega}=10.24$ ms and 40.96 ms (Fig.~\ref{fig:4}b-2 and \ref{fig:4}b-3) have the first peaks around 7 ms and 25 ms, respectively, leading to non-zero $P_{1}$.
$P_{1}(T_{\Omega})$ calculated for the data $p_{fold}(t)$ of our 1D simulation for P5GA hairpin (filled circles in Fig.~\ref{fig:4}c) is non-monotonic with $T_{\Omega}$ and is maximized around $T_{\Omega}=T^{SR}\approx \tau_F^o+\tau_U^o\approx 10.8$ ms.
Notably, $P_1(T_{\Omega}$) rise sharply for $T_{\Omega}\gtrsim T^{RA}\simeq 2.56$ ms (Fig.~\ref{fig:4}c) and the $T^{SR}$ that maximizes $P_1$ differs from the $T^{RA}$ that minimizes $\tau_F$.
The $P_{1}$ for the unfolding with $f=13$ pN and $\delta f=1.4$ pN, by calculating unfolding time distribution $p_{unfold}(t)$, are also depicted with empty circles in Fig.~\ref{fig:4}d. We find that the unfolding is also synchronized with an optimal driving period around $T_{\Omega}\approx \tau_F^o+\tau_U^o\approx 8$ ms (see Fig.~\ref{fig:1}c).
Thus, RNA folding (unfolding) becomes maximally coherent at  $T^{SR}\approx \tau_F^o+\tau_U^o$.
Due to asymmetric shape of potential leading to $\tau_F\gg \tau_U$, the SR condition for folding becomes $T_{\Omega}\approx\tau_F^o+\tau_U^o\approx \tau_F^o$, which is worth comparing with the conventional SR condition $T_{\Omega}\approx\tau_F^o+\tau_U^o = 2\tau_F^o$ for symmetric bistable systems.

There are many alternative measures to demonstrate the manifestation of SR.
One measure, directly amenable to experiment, is the power spectrum $S(\nu)=\int_{-\infty}^{\infty}e^{-i 2\pi \nu \tau}\langle z(t+\tau)z(t)\rangle d\tau$ where $\langle\ldots\rangle$ denotes ensemble and long time averages for the trajectory $z(t)$ \cite{gammaitoni1998stochastic}.
The $S(\nu)$ shows a narrow sharp peak with the frequency $\nu=1/T_{\Omega}\approx 1/10$ ms$^{-1}$ (see the inset of Fig.~\ref{fig:4}b-2).
In addition, spectral behavior of dynamic susceptibility defined $\chi(\nu)=\langle z(\nu)\rangle/f(\nu)$, which also becomes optimal at $\nu\approx 1/10$ ms$^{-1}$, can be used to quantify the response of system to an external driving force.\\

{\bf Biopolymers can filter the resonance frequency from the spectrum of an external signal :}
Due to the multitude of cellular processes nonequilibrium noise in the cell is decomposed into various frequencies.
To check if our model can discriminate the resonant frequency using SR,
we performed the 1D simulation for RNA hopping transitions under $f=17$ pN by adding oscillatory forces consisting of different angular frequencies $\sum_{i=1}^{3}\delta f_{i} \sin(\Omega_i t)$ where $\Omega_{1}=2 \pi/0.1 $ ms$^{-1}$, $\Omega_{2}=2 \pi/10 $ ms$^{-1}$ and $\Omega_{3}=2 \pi/100 $ ms$^{-1}$; here, the optimal folding period for the SR, $T_{SR}\approx 10$ ms, is included in the $i=2$.
The $p_{fold}(t)$ for uniform amplitudes of the additive forces with $\delta f_{1}=\delta f_{2}=\delta f_{3}=1.4$ pN (Fig.~\ref{fig:5}) displays an oscillatory pattern with multiple peaks, suggesting that the folding is synchronous with the oscillatory force with the period $T^{SR}$.
Even for the case of nonuniform amplitudes, in which $\delta f_{1}$ and $\delta f_{3}$ have greater values than $\delta f_2$, the signature of synchronization to the oscillatory driving with $T^{SR}$ is retained (Fig.~\ref{fig:5}).
Furthermore, even when we choose $\delta f$ randomly from the values distributed between 0 and 2 pN at every time setp, we still found the $p_{fold}(t)$ with multiple peaks, which suggests that the P5GA system is capable of filtering the signal with frequency of $2 \pi/T^{SR}$ out of the nonequilibrium noise.\\

{\bf Experimental demonstration of SR in biomolecular dynamics :} 
Although the results discussed above are based on the numerical simulation that applies mechanical force directly acting to the ends of RNA hairpin without mediating polymer handle as in actual optical tweezers experiment, our findings on SR and RA conditions for folding transition should not change qualitatively. As quantitatively discussed by Hyeon \emph{et al.} \cite{Hyeon08PNAS} as well as recognized by others \cite{Bustamante01Sci}, the presence of handle polymers simply slows down the transition kinetics by pinning the motion of the two end points of biopolymers connected to the handles. Furthermore, in all practical situations, the conformational transition time of biopolymers is much slower than the tension propagation time \cite{Manosas07BJ}. Therefore, even in the presence of handles SR is still realized at the optimal frequency, corresponding to the sum of the folding and unfolding transition times that are measured in the presence of handles. 
Many-body simulation using SOP model of RNA hairpin in the presence of 20 nm handles attached to the two ends of hairpin confirms the realization of SR in the presence of handles (see \textit{SI text} and Fig.S1). 
For the purpose of experimental demonstration of SR using biopolymers, it is recommended that the experiment is conducted at the transition mid-force ($f_m$), at which the folding and unfolding transitions are the most frequently observed \cite{Bustamante01Sci,Block06Science} (Fig.S2). In this case, SR condition
is expected to be formed around twice the mean hopping
transition time ($\approx 2\tau_F(f=f_m)$).

\section*{CONCLUDING REMARKS}
Originally, the SR refers to the phenomenological observation that
a system under an optimal thermal noise can amplify a weak external signal if the system is nonlinear.
This paper restates the SR and RA condition by focusing more on the transition dynamics of biopolymers rather than the amplification of signal. 
Furthermore, unlike the many studies that consider the SR using noise strength, i.e., temperature, as a variable, the present study focuses on the ``bona-fide SR" with respect to the driving frequency. For biologically relevant environment, whose temperature range should be narrowly defined, it is more appropriate to consider the effect of driving frequency on SR. 
Folding and unfolding transitions of biopolymers enhanced under and synchronized to an external periodic signals, which is clearly demonstrated in our simulations, suggest that as long as the SR condition is met, conformational sampling of biopolymers in vivo occurs more frequently than in vitro, in response to the spectrum of athermal noise in the cell.    

True manifestation of SR and RA in the context of biopolymer dynamics in vivo should hinge on existence of the actual  periodic frequency that is in resonance to the transition dynamics of biopolymer.
It would be of great interest to investigate the power spectra of cellular noise and how biological dynamics, for instance, conformational transitions of riboswitch, are modulated in the presence of athermal signals from other dynamic components such as motion of ribosome translocation and associated gene regulation \cite{Greenleaf08Science}.



\section*{MATERIALS and METHODS}
{\bf Model :} We used the SOP model to simulate the dynamics of P5GA RNA hairpin under tension. In SOP model, the effective Hamiltonian is defined as follows \cite{HyeonBJ07,Hyeon08PNAS}:
\begin{align}
&H_{SOP}=-\frac{k R_{0}^{2}}{2}\sum_{i=1}^{N-1}\ln\left[1-\frac{(r_{i,i+1}-r_{i,i+1}^{o})^2}{R_{0}^{2}}\right]\nonumber\\
&+\sum_{i=1}^{N-3}\sum_{j=i+3}^{N}\epsilon_{h}\left[\left(\frac{r_{i,j}^{o}}{r_{i,j}}\right)^{12}-2\left(\frac{r_{i,j}^{o}}{r_{i,j}}\right)^{6}\right]\Delta_{i,j}\nonumber\\
&+\sum_{i=1}^{N-3}\sum_{j=i+3}^{N}\epsilon_{l}\left(\frac{\sigma}{r_{i,j}}\right)^{6}(1-\Delta_{i,j})+\sum_{i=1}^{N-2}\epsilon_{l}\left(\frac{\sigma^{\ast}}{r_{i,i+2}}\right)^{6},
\label{eq:sop}
\end{align}
where we modeled each nucleotide as an interaction center; thus $N=22$.
The first term restrains the bond length between the neighboring beads by using the finite extensible nonlinear elastic potential with $k=20$ kcal/(mol$\times$\AA$^2$), and $R_0=0.2$ nm.
Here, $r_{i,i+1}$ is the distance between neighboring interaction centers $i$ and $i+1$, and $r_{i,i+1}^o$ is the distance in the native structure.
Lennard-Jones potential was employed to stabilize the base pair present in the native structure in the second term with $\epsilon_h=0.7$ kcal/mol, but make the non-native base pair repulsive in the third term, so that the model can easily reach the native topology without being trapped into non-native conformations.
If $i$ and $j$ are in contact in the native state $\Delta_{ij}=1$; otherwise $\Delta_{ij}=0$.
$r_{i,j}^o$ is the distance between $i$ and $j$ interaction centers that satisfy $|i-j|>2$ and $\Delta_{ij}=1$. For non-native pairs, the strength and range of interaction were set by $\epsilon_l=1$ kcal/mol and $\sigma=0.7$ nm.
The last term involving the repulsion with $\sigma^*=0.35$ nm maintains a certain angle between $i$ and $i+2$ interaction centers.\\

\begin{figure*}
\centering
\includegraphics[width=16cm]{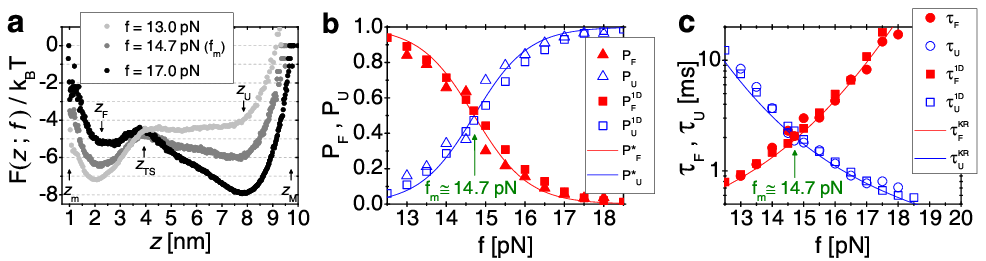}\caption{\label{fig:1} Thermodynamics and kinetics of P5GA hairpin under tension. (a) Free energy profile as a function of molecular extension $z$ under tension $f$, obtained from the molecular simulation using SOP model.
Shown are the three free energy profiles at $f=13.0$, $14.7$, and $17.0$ pN.
At $f=f_m=14.7$ pN, the probability of being  in UBA and NBA are equal.
The profile is tilted towards UBA (or NBA) when $f=17.0>f_m$ (or $f=13.0<f_m$).
(b) Unfolding and folding probability $P_U(f)$ and $P_F(f)(=1-P_U(f))$ of P5GA against tension $f$.
Results from the SOP model ($P_{F}(f)$ and $P_{U}(f)$) and two state model (solid curves) are in good agreements.
$P^{1D}_{U}(f)$ (or $P^{1D}_F(f)$) calculated from the tilted 1D free energy profile interpolated from results of the SOP model displays a good agreement with other results.
(c) Mean folding time ($\tau_{F}(f)$) and unfolding time ($\tau_{U}(f)$) at varying tensions, obtained from the SOP simulation (circles),
$\tau_{F}^{1D}(f)$ and $\tau_{U}^{1D}(f)$ obtained from the 1D simulation on the tilted free energy profile (squares), and
mean first passage times $(\tau_{F}^{KR},~\tau_{U}^{KR})$ calculated using the tilted free energy profile (solid curves) are in a good agreement.
The transition mid-force determined at the intersection between $\tau_F(f)$ and $\tau_U(f)$ is $f_{m}=14.7$ pN.}
\end{figure*}

\begin{figure*}
\centering
\includegraphics[width=16cm]{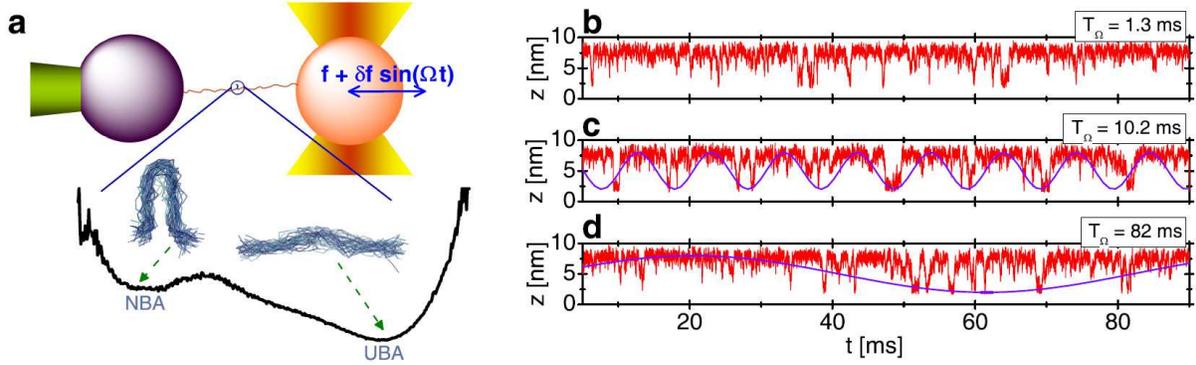}
\caption{\label{fig:2} Demonstration of SR using simulation of P5GA hairpin under an oscillating force.
(a) Structural ensemble of P5GA hairpin at NBA and UBA shown on the free energy profile $F(z;f=17$ pN$)$.
Time trajectories of molecular extension $z(t)$ at constant tension $f=17$ pN are shown from the simulation of SOP model when a small time dependent oscillatory force $\delta f\sin{(2\pi t/T_{\Omega})}$ ($\delta f = 1.4$ pN) is added with (b) $T_{\Omega}(=2\pi/\Omega)=1.3$ ms, (c) $10.2$ ms, and (d) $82$ ms. Coherent hopping transition between folding and unfolding is found with $T_{\Omega}=T^{SR}\approx10.2$ ms, which clearly shows the SR.
The oscillatory forces with $T_{\Omega}=10.2$ ms and $82$ ms are overlaid on the time trajectory in c and d, respectively, to emphasize the effect of synchronization. See Fig.S1 for the SR of P5GA hairpin in the presence of handles.}
\end{figure*}

\begin{figure}
\centering
\includegraphics[width=16cm]{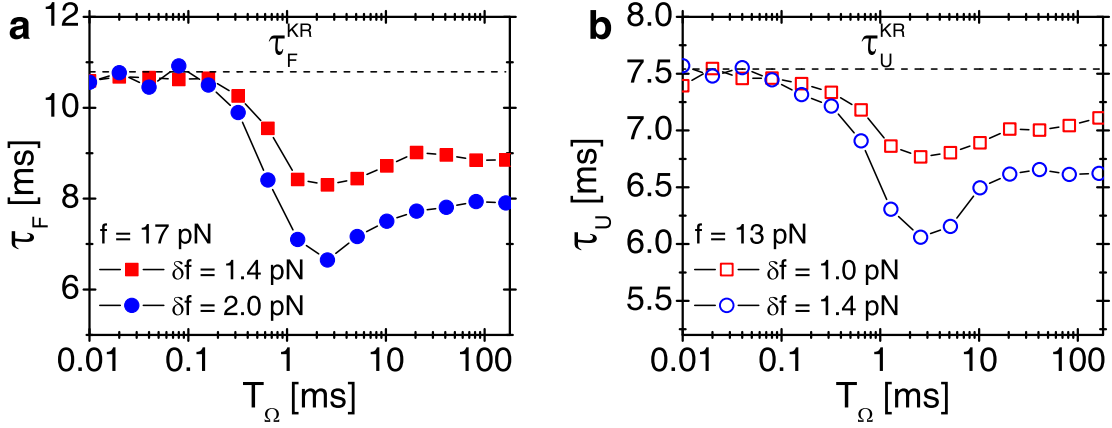}
\caption{
\label{fig:3} Resonant activation (RA) in folding and unfolding transitions of P5GA. (a) Mean folding time $\tau_{F}$ of P5GA under $f=17$ pN as a function of $T_{\Omega}$ with $\delta f=1.4$ pN (filled squares) and $\delta f=2.0$ pN (filled circles). The transition times are minimized at $T_{\Omega}=T^{RA}\simeq 2.56$ ms. (b) Mean unfolding time $\tau_{U}$ of P5GA under $f=13$ pN as a function of $T_{\Omega}$ with $\delta f=1.0$ pN (empty squares) and $\delta f=1.4$ pN (empty circles).
The transition times are minimized at $T_{\Omega}=T^{RA}\simeq 2.56$ ms.}
\end{figure}

\begin{figure}
\centering
\includegraphics[width=12cm]{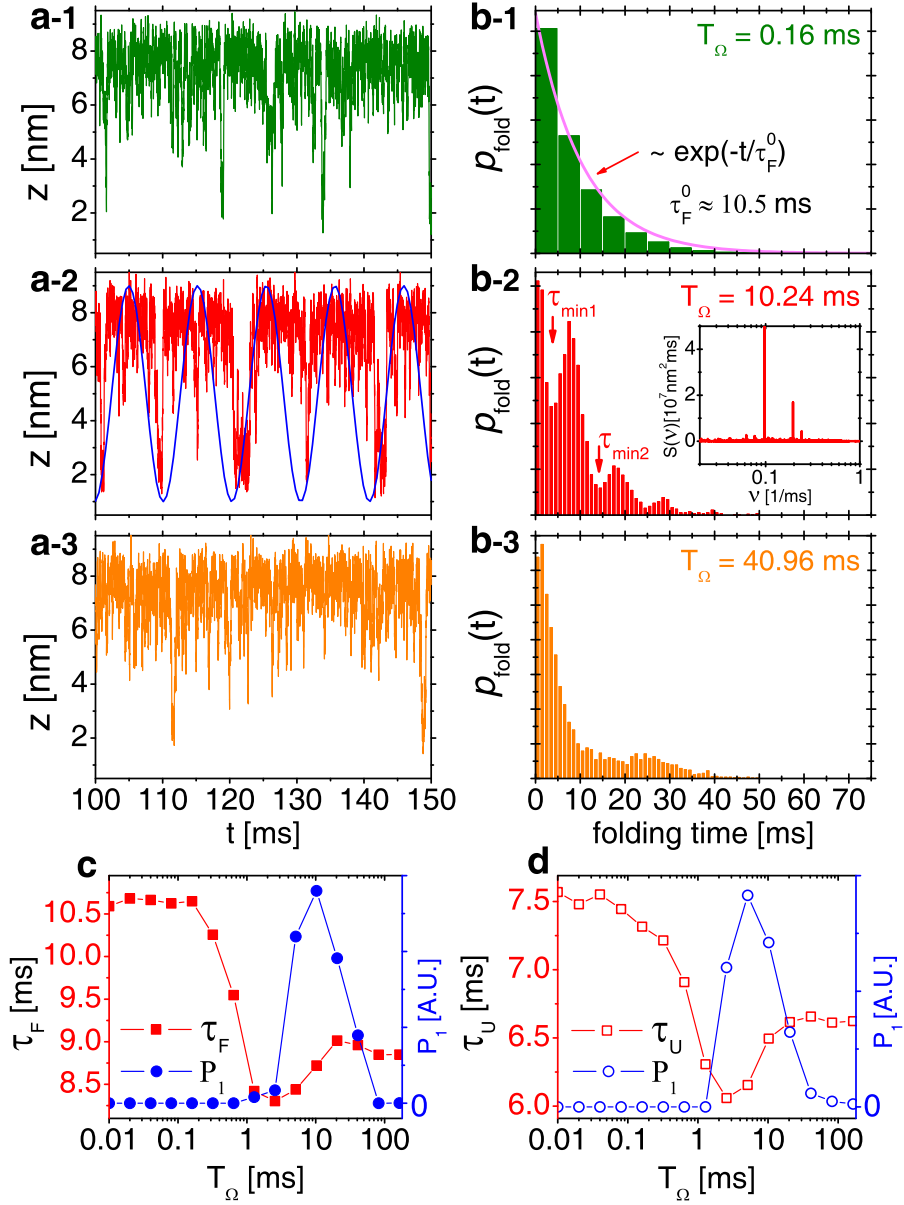}
\caption{
\label{fig:4} SR in the folding and unfolding transitions of P5GA under time dependent tension $f+\delta f\sin(\Omega t)$. (a) Time trajectories and (b) folding time distributions $p_{fold}(t)$ with $f=17$ pN and $\delta f=1.4$ pN at $T_{\Omega}=0.16$ ms, $10.24$ ms, and $40.96$ ms are shown. The synchronization (SR condition) occurs around $T_{\Omega}=T^{SR}\approx\tau_F^o+\tau_U^o\approx10.8$ ms, as shown in (a-2). The inset in (b-2) depicts the power spectrum $S(\nu)$, showing the sharpest peak at $\nu=1/T^{SR}\approx 0.1$ ms$^{-1}$. (c) Mean folding time $\tau_{F}$ and measure of coherence $P_1$ as a function of $T_{\Omega}$.
(d) Mean unfolding time $\tau_{U}$ and measure of coherence $P_{1}$ as a function of $T_{\Omega}$ with $f=13$ pN and $\delta f=1.4$ pN.
Note that the SR and RA conditions are made at different $T_{\Omega}$s in (c) and (d). }
\end{figure}

\begin{figure}
\centering
\includegraphics[width=16cm]{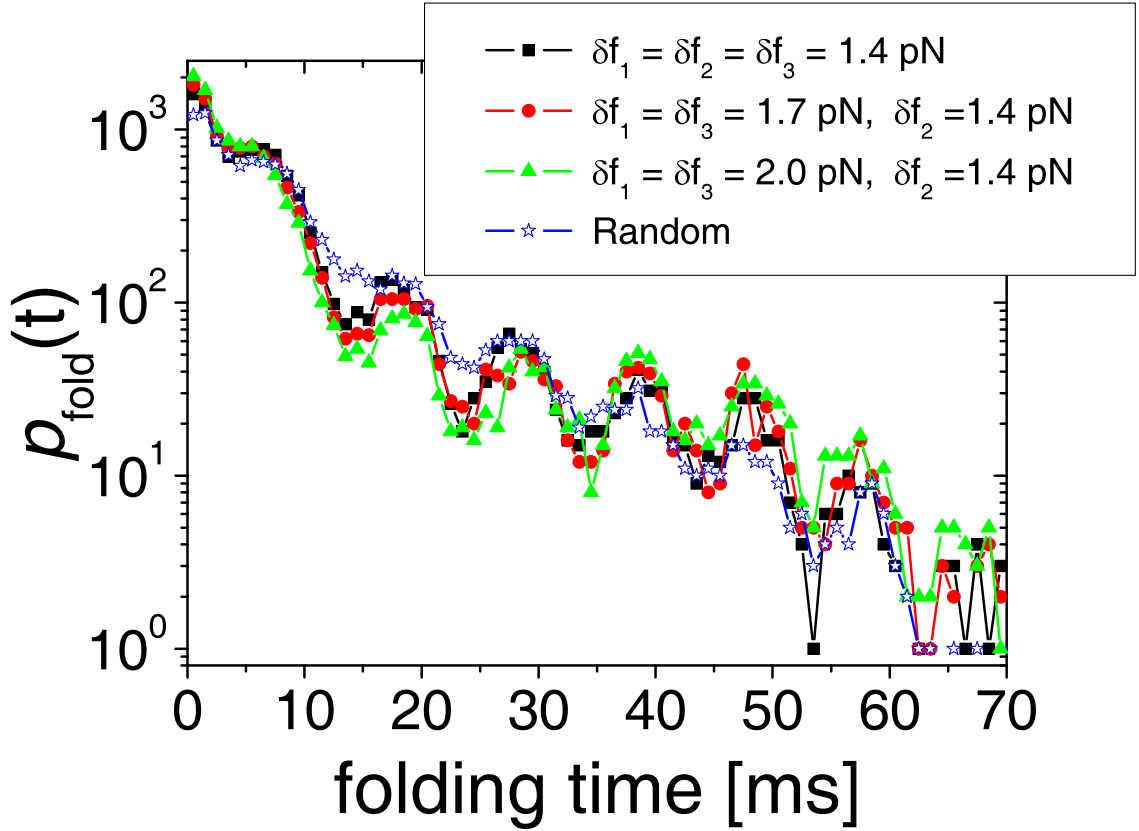}
\caption{\label{fig:5}
Filtering of SR condition.
Shown are the $p_{fold}(t)$ of P5GA under $f+\sum_{i=1}^{3}\delta f_{i} \sin(\Omega_i t)$ at varying combinations of $\delta f_i$ with $\Omega_{1}=2 \pi/0.1 $ ms$^{-1}$, $\Omega_{2}=2 \pi/10 $ ms$^{-1}$, $\Omega_{3}=2 \pi/100 $ ms$^{-1}$ and $f=17$ pN.
For filled symbols, we fix $\delta f_{2}=1.4$ pN, corresponding to the amplitude for the optimal frequency $\Omega_{2}$ for SR, and vary the other amplitudes to be $\delta f_{1}=\delta f_{3}=1.4$ pN (filled squares), 1.7 pN (filled circles) and 2.0 pN (filled triangles), respectively.
The starred symbols represent the cases where $\delta f_{1}$, $\delta f_{2}$ and $\delta f_{3}$ are taken randomly between 0 and 2 pN every time step to make a nonequilibrium noise $\delta f(t)$. Notably, the folding transition filters the optimal driving period despite other drivings with larger or transient amplitudes.}
\end{figure}

{\bf Simulation :} We performed Brownian dynamics simulation by numerically integrating the discretized equation of motion
\begin{align}
&{\vec{r}}_{\alpha,i}(t+\Delta t)=\vec{r}_{\alpha,i}(t)+\frac{\Delta t}{\zeta}\left(-\frac{\partial H_{SOP}}{\partial r_{\alpha,i}}+f\delta_{i,22}\hat{e}_z+\vec{\xi}_{\alpha,i}(t)\right)\nonumber\\
&\vec{r}_{i=1}(t)=0
\end{align}
where $\alpha=x,y,z$ and the random force $\vec{\xi}_{\alpha,i}(t)$ satisfies $\langle\vec{\xi}_{\alpha,i}(t)\rangle=0$ and
it is assumed that the probability distribution of the noise in the random force is Gaussian, i.e.,
$\mathcal{P}[\vec{\xi}_{\alpha,i}(t)]\propto\exp\left(-\int d\tau\frac{[\xi_{\alpha,i}(\tau)]^2}{4\zeta k_BT}\right)$ and
$\langle\vec{\xi}_{\alpha,i}(t)\cdot \vec{\xi}_{\beta,j}(t+n\Delta t)\rangle=\frac{2\zeta k_{B}T}{\Delta t}\delta_{0,n}\delta_{i,j}\delta_{\alpha,\beta}$ with $n=0,1,2,...$.
We probed the folding-unfolding dynamics of RNA hairpin by applying tension $f$ to the 3'-end of the hairpin while fixing the position of 5'-end at temperature $T=300$ K.
Also, we chose the friction coefficient $\zeta=100 m/\tau_L$ with $\tau_L=(ma^2/\epsilon_h)^{1/2}\approx 3-5$ ps corresponding to the value that a nucleotide whose size $a\approx 0.5$ nm feels in water environment whose viscosity is $\sim 1$ cP \cite{HyeonBJ07}.
We applied the tension $f$ between $13\leq f\leq18$ pN (Fig.~\ref{fig:1}) to the molecule and added small sinusoidal forces of $\delta f \sin{(\Omega t)}$ with varying $\Omega$ values (Fig.~\ref{fig:2}). We set the simulation time step $\Delta t = 4.7\times10^{-5}~\mu$sec and recorded data at every $10^{5}$ time steps ($4.7~\mu$sec).
After equilibrating the RNA in native state for $\sim 47$ $\mu s$, we applied tension $f$ or $f+\delta f\sin{(\Omega t)}$ and recorded the $z(t)$ of each trajectory for $t > 10^{2}$ ms.


\begin{acknowledgments}
This work was supported by Brain-Korea 21 program and National Core Research Center for Systems Bio-Dynamics to W.S. and National Research Foundation of Korea grant (2010-0000602) to C.H. We thank Korea Institute for Advanced Study for providing computing resources.
\end{acknowledgments}

\clearpage

\clearpage

\clearpage

\clearpage

\clearpage

\clearpage

\section*{SUPPORTING INFORMATION (SI)}

{\bf Stochastic resonance in the presence of handles :}
To study whether the SR is still realized when the weak signal is indirectly exerted to the system through handle polymers, just like in the real experimental condition in laser optical tweezers (LOTs), we simulate folding dynamics of an RNA hairpin by attaching stiff handles to the two of the P5GA RNA hairpin modeled with SOP representation (Eq.[2] in the main text). 
Each handle is modeled by using the following energy potential \cite{Hyeon08PNAS_S}:  
\begin{equation}
H_{\text{handles}}=\frac{k_S}{2}\sum_{i=1}^{N-1}({r_{i,i+1}}-{r_0})^2-{k_A}\sum_{i=1}^{N-2}{{\bf{\hat{r}}}_{i,i+1}}\cdot{{\bf{\hat{r}}}_{i+1,i+2}}.
\label{eq:s1}
\end{equation}
The first term denotes the chain connectivity, with an equilibrium distance ${r_0}=0.5$ nm and spring constant ${k_S}= 1.4\times 10^{4}~\text{pN}\cdot\text{nm}^{-1}$. The second term is the bending penalty potential with bending modulus $k_A=561$ pN$\cdot$nm, which gives rise to the persistence length ${l_p}=70$ nm. We take $N=40$ for the length of handle, which amounts to the contour length of $L=20$ nm.
Fig.~\ref{fig:s1} shows time trajectories of the extension of the entire system, which includes the extension of two handles, at $f=17$ pN and $\delta f=1.4$ pN, for three different driving periods. The molecular extension $z(t)$ (left) and the total extension $z_{L}(t)$ (right) are depicted in the red and black axes, respectively.  
As shown in the trajectories, the phase of $z_{L}(t)$ is compatible with that of $z(t)$. 
Even in the presence of handles, the SR is realized at an optimal period $60$ ms (Fig.~\ref{fig:s1}b) where the folding transitions are synchronized to the frequency of external driving.  
The other trajectories under relatively fast (Fig.~\ref{fig:s1}a) and slow (Fig.~\ref{fig:s1}c) driving show random transitions regardless of the drivings. 
As it is discussed in the main text and in the Ref.\cite{Hyeon08PNAS_S}, 
the presence of the handles pins the overall molecular motion and slows down the transition kinetics of the molecule sandwiched between the handles. 
We have estimated that the transition dynamics with handles with $l_{p}=70$ nm and $L=20$ nm tend to be almost about six-fold slower \cite{Hyeon08PNAS_S} than the Kramers times ($\approx 10$ ms for folding and $\approx0.7$ ms for unfolding) without handles which we have considered in the text, and found SR (Fig.~\ref{fig:s1}b) at an optimal driving period $T_{\Omega}\approx 60$ ms which is approximately sum of the Kramers times.\\

\noindent{\bf Stochatstic resonance at the the transition mid-force :}
To demonstrate the SR using LOTs experiment, it would be most straightforward to perform the experiment around the external force value near the transition mid-force, at which the transition from  one basin to another is most frequent.  
To show SR under the transition mid-force ($f_{m}=14.7$ pN), we have performed the SOP simulations for hopping (folding and unfolding) dynamics of a P5GA under $f_{m}$. 
Fig.~\ref{fig:s2} shows time trajectories of the P5GA extension $z(t)$ for three different driving periods, at $\delta f=1.4$ pN. 
SR is clearly shown in the Fig.~\ref{fig:s2}b at an optimal driving period $T_{\Omega}=4$ ms. In this case, the optimal driving period is found to be around twice the mean hopping transition timee ($\approx 2\tau_F(f=f_m)\approx 4$ ms, see also Fig.1c where $\tau_F=\tau_U=$2 ms), in consistent with the \emph{time-scale matching condition} as has extensively been discussed in the SR community \cite{gammaitoni1998stochastic_S,srra_S,bonafide_S}. 
The other trajectories, under relatively faster (Fig.~\ref{fig:s2}a) and slower (Fig.~\ref{fig:s2}c) drivings which deviate from the optimal frequency, show random transitions.

\renewcommand{\thefigure}{S1}
\begin{figure*}
\centering
\includegraphics[width=14cm]{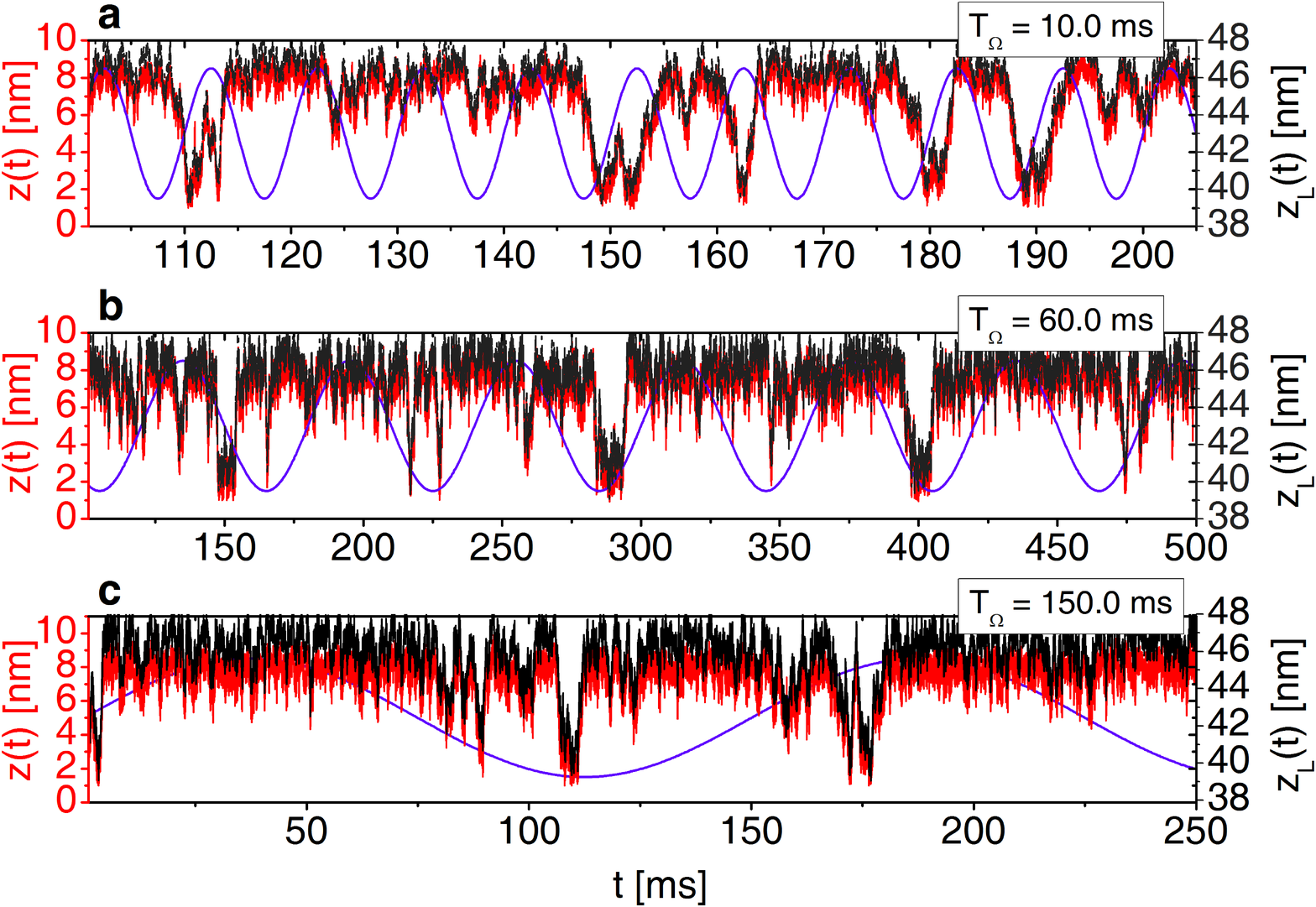}
\caption{\label{fig:s1} Demonstration of RNA hairpin trajectories under SR condition in the presence of handles. Together with the SOP representation of RNA hairpin, we considered the handle of contour length $L=20$ nm and persistence length $l_{p}=70$ nm. 
The time trajectories of molecular extension $z(t)$ (red;left axis) and the extension with handles $z_{L}(t)$ (black;right axis) are displayed at constant tension $f=17$ pN, driving force $\delta f \sin(\Omega t)$ with $\delta f=1.4$ pN, and period $T_{\Omega}=2\pi/\Omega$.
The phase of extensions $z_{L}(t)$ overlaid on the $z(t)$ is in good agreements. Both $z(t)$ and  $z_{L}(t)$ show random transitions at (a) $T_{\Omega}=10.0$ ms and (c) $T_{\Omega}=150.0$ ms.
(b) The folding transitions are synchronized with an optimal driving period $T_{\Omega}=60.0$ ms. The blue curves are overlaid to show the phase of oscillatory driving forces.}
\end{figure*}
\clearpage

\renewcommand{\thefigure}{S2}
\begin{figure*}
\centering
\includegraphics[width=14cm]{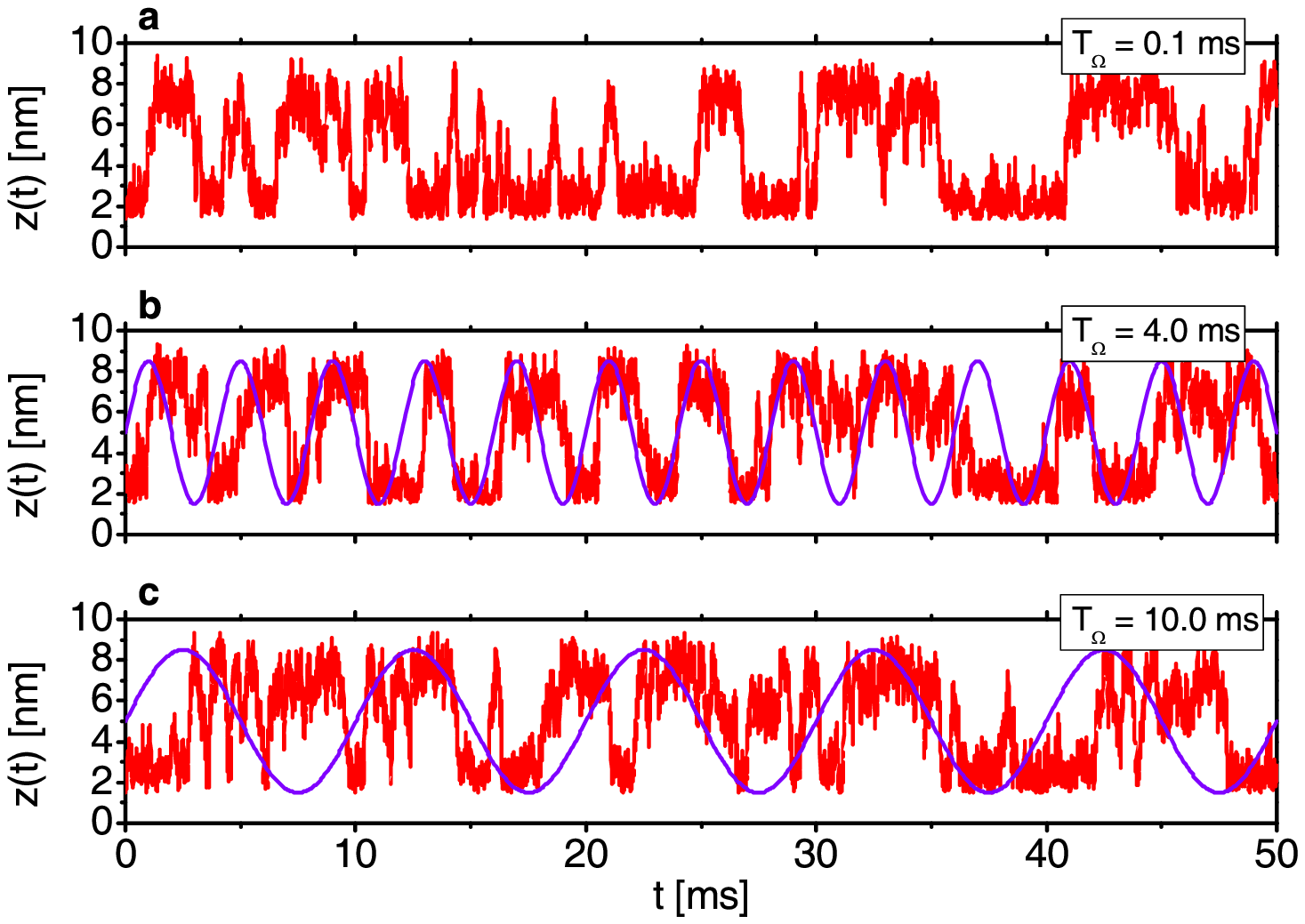}
\caption{\label{fig:s2} Demonstration of bona-fide SR using the SOP model under the transition mid-force $f_{m}=14.7$ pN in the absence of handles. We depicted time trajectories of molecular extension $z(t)$ at $f_m$ and $\delta f \sin(\Omega t)$ with $\delta f=1.4$ pN and period $T_{\Omega}=2\pi/\Omega$.
$z(t)$ show random transitions at (a) $T_{\Omega}=0.1$ ms and (c) $T_{\Omega}=10.0$ ms, while coherent transitions occur at an optimal period (b) $T_{\Omega}=4.0$ ms. 
The blue curves are overlaid to show the phase of oscillatory driving forces.}
\end{figure*}

\end{document}